\RequirePackage{lineno}

\documentclass[aps,prb,twocolumn,showpacs]{revtex4}
\usepackage{hyperref}
\usepackage{graphicx}
\DeclareGraphicsExtensions{.png,.jpg,.eps,.pdf}

\usepackage{xcolor}
\usepackage{amsmath}

\begin{document}

\title{Exchange rules for diradical $\pi$-conjugated  hydrocarbons}

\author{R. Ortiz$^{1,2,4}$, R. A. Boto$^{3}$, N. Garc\'ia-Mart\'inez$^{1,2}$, J. C. Sancho-Garc\'ia$^{4}$, M. Melle-Franco$^{3}$, 
J. Fern\'andez-Rossier$^{1}$\footnote{On leave from Departamento de F\'isica Aplicada, Universidad de Alicante, Spain} }
\affiliation{(1) QuantaLab, International Iberian Nanotechnology Laboratory (INL), Av. Mestre Jos\'e Veiga, 4715-330 Braga, Portugal
}

\affiliation{(2)Departamento de F\'{\i}sica Aplicada, Universidad de Alicante, 03690,  Sant Vicent del Raspeig, Spain
}

\affiliation{(3) CICECO, Departamento de Qu\'{\i}mica, Universidade de Aveiro, 3810-193 Aveiro, Portugal
}

\affiliation{(4)Departamento de Qu\'{\i}mica F\'{\i}sica, Universidad de Alicante, 03690, Sant Vicent del Raspeig, Spain
}

\date{\today}

\begin{abstract}

A variety of planar $\pi$-conjugated  hydrocarbons such as heptauthrene, Clar's goblet and, recently 
synthesized, triangulene have two electrons occupying two degenerate molecular orbitals. The resulting spin of the interacting 
ground state is often  correctly anticipated as $S=1$, extending the application of Hund's rules to these systems, but  this is not correct in some instances.  
Here we provide a set of rules to correctly predict the existence of zero mode states, as well as the spin multiplicity of both the ground state and   the low-lying excited states, together with 
their  open- or closed-shell nature. This is accomplished using a combination of analytical arguments and configuration interaction calculations with a 
Hubbard model,    both backed by quantum chemistry  methods with a larger Gaussian  basis  set. Our results 
 go beyond the well established Lieb's theorem and Ovchinnikov's rule, as we address the  multiplicity and the open-/closed-shell nature of both 
ground and excited states.

\end{abstract}

\maketitle

\section{Introduction}
Understanding the so-called  exchange  interactions in atoms, molecules and crystals is one of the central  topics in the study of the electronic 
properties of matter. In the case of atoms, whether or not a given atom has a magnetic ground state is a long settled issue: only open-shell atoms can 
have a magnetic moment whose magnitude is established by the so-called Hund's rules. The nature of inter-atomic exchange is also well understood in a wide 
class of oxides, since the seminal work of Goodenough\cite{goodenough63} and Kanamori\cite{kanamori59}.  

The study of $\pi$-conjugated  hydrocarbons with diradical nature (which also includes polycyclic aromatic hydrocarbons (PAHs) or nanographenes) 
goes back to more than a century ago\cite{Schlenk,Borden77,rajcareview,moritaReview}.  Experimental research in this area has faced the challenge of the very high chemical reactivity of diradicals. However,  recent developments of new synthesis 
 methods,  in ultrahigh vacuum surfaces,   has made it possible to synthetize   highly reactive diradical species, such as triangulene\cite{pavlivcek2017} or graphene  nanoribbons with zigzag edges\cite{wang2016}  as well as other open-shell PAHs \cite{NachoNat}, and to explore them   using scanning probe microscopies. 
 
 Diradicals host two degenerate, or almost degenerate, states that lie in the gap between the doubly occupied and empty molecular orbitals that are formed primarily  by $\pi$ orbitals\cite{Yoneda2009}. Very often they are localized in one of the two interpenetrating triangular sublattices that form the honeycomb bipartite lattice.  
Naively, one could expect that diradicals always had a ground state with $S=1$, due to the similitude with the case of open-shell atoms. Whereas this is 
often the case in a variety of systems \cite{melle2015, Schlenk, gryn'ova2015, Lischka2016, Sandoval2019, Yoneda2011}, in others there is a {\em violation} of the Hund's rule\cite{Yazyev09,gryn'ova2015, Sheng2013, Trinquier2015, Malrieu2016, Trinquier2018, Pogodin1985}.  Therefore,  the sign of the exchange interaction in this class of diradicals is  not always the same\cite{rajcareview}.   

In general, the spin of the ground state in PAHs can be anticipated using the Ovchinnikov's rule\cite{ovchinnikov1978}, that states that the spin $S$ of the ground state is given by $S=\frac{N_A-N_B}{2}$, where $N_{A,B}$ are the number of C atoms in each of the interpenetrating triangular sublattices that form the honeycomb lattice.   Interestingly, Lieb upgraded  the Ovchinnikov's rule  into a theorem\cite{Lieb89},  which states that the exact interacting ground state of the Hubbard model for a bipartite system is given also by $S=\frac{N_A-N_B}{2}$.   Yet another theorem\cite{sutherland86} establishes that for a bipartite lattice the number of zero modes is given by $N_Z=|N_A-N_B|$. In consequence, systems with a sublattice imbalance of $N_A=N_B\pm 2$, such as triangulene,  are diradicals with $S=1$.

However, this picture is not complete for several reasons. First, some diradicals, such as the Clar's goblet or the  undistorted (planar)  cyclobutadiene have no sublattice imbalance and still have two zero modes and, in agreement with the Ovchinnikov's rule and Lieb's theorem, have  $S=0$.   Second,  whereas Lieb's theorem\cite{Lieb89} has predictive power on the spin $S$ of the ground state, it has a very poor explaining power: what is the nature of the antiferromagnetic exchange in those diradicals with a GS with $S=0$?   Third, the Lieb's theorem does not provide any information on spin of the excited states,  nor on their open-shell vs closed-shell nature.

In this work  we address  these longstanding research questions in the case of planar conjugated hydrocarbons or  nanographenes (as well as its generalization to all kind of diradical $\pi$-conjugated hydrocarbons in the suppl. mat.).  
First,  we provide a unified approach to anticipate if a given bipartite system has zero mode states. This constitutes a prerequisite for open-shell configurations.  Second,  
 we provide a  set of rules  that determine the multiplicity and the open-/closed- shell nature of the lowest in energy many-body states for these nanographene diradicals.  We find that the key ingredient that defines the properties in the multi-electronic problem are the transformation properties of the molecular orbitals under the symmetry operations of the point group of the molecule.  We therefore establish a set of simple rules that permit to anticipate the relative position and approximate excitation energy  of the six lowest energy multi-electron states,  a  degenerate triplet and three singlets.

 \begin{figure}
 \centering
    \includegraphics[width=0.5\textwidth]{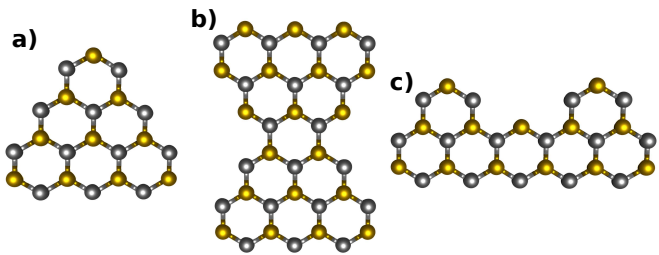}
\caption{ Atomic structure of a) triangulene, b) Clar's goblet and c) heptauthrene. We assign silver and gold color to highlight  the triangular sublattices. }
\label{sublattices}
\end{figure}
 
     Our results are based on an analysis using theory  at three  levels of complexity. First, an analytical description of the Hubbard model for these compounds when the active space is restricted to the two in-gap states only. Second, numerical calculations, still within the Hubbard model, in a larger active space that includes occupied and virtual molecular orbitals. Third,  ab initio quantum chemistry methods\cite{cusinato2018}, including  complete active space (CAS) configuration interaction (CI)  calculations, carried  using a gaussian orbital basis, and further corrected by second-order N-electron Valence Perturbation Theory (NEVPT2).  
   
\section{ Zero modes (non-bonding states).}
We first revisit the  problem\cite{longuet50,Borden77} of predicting the existence of non-bonding zero modes in a PAH. Our discussion applies when the system is described with a tight-binding model with one $p_z$ orbital per atom in the first neighbor hopping approximation.  This defines a tight-binding model in a bipartite graph and permits to  provide a unified  and compact picture that accounts for a number of results derived over the years\cite{sutherland86,inui1994,Fajtlowicz2005OnMM,Fernandez2007, weikprb2016}. Bipartite graphs  can be drawn as the superposition of two interpenetrating sublattices  (see Fig.1).  We define the $\Gamma$ matrix that assigns a $+1$ to the carbon sites of the $A$ sublattice and a $-1$ to the carbon sites of the $B$ sublattice. 

The H\"uckel (tight-binding) Hamiltonian ${\cal H}_0$  that describes the $\pi$ orbitals has  the  so-called chiral symmetry\cite{Ryu02,delplace2011},  given by the equation: 
\begin{equation}
{\cal H}_0\Gamma+ \Gamma {\cal H}_0=0
\label{anticom}
\end{equation}   
This anti-commutation relation,  different from usual  commutators associated to symmetries, 
entails several consequences relevant for the ensuing discussion.  First, let us consider the eigenstates ${\cal H}_0 |\psi_n\rangle=E_n|\psi_n\rangle$.  Eq.(\ref{anticom}) implies that 
$|\psi'_n\rangle \equiv \Gamma |\psi_n\rangle$
 is also an  eigenstate of ${\cal H}_0$ with eigenvalue $-E_n$.
  If  $E_n\neq 0$,  $|\psi_n\rangle$ and $|\psi'_n\rangle$ have to be orthogonal, as they are eigenstates with different eigenvalues.  Their orthogonality can be
 written up as:
 \begin{equation}
\langle \psi_n |\Gamma|\psi_n\rangle=0 \;\;\;  \left( E_n\neq 0\right)
\end{equation}
which implies an equal weight on the two sublattices. 

 Hamiltonians ${\cal H}_0$  with chiral symmetry can also host states  with $E_n=0$, that we label as $\phi_\eta$.   It can be easily  demonstrated  (see suppl. mat.)  that these zero modes are sublattice polarized  and,  therefore,  are eigenstates of $\Gamma$. As a result,  they  satisfy $\langle \phi_\eta |\Gamma |\phi_\eta\rangle=\pm1$, depending on whether they are localized in the $A$ or $B$ sublattice. Thus, Eq.(\ref{anticom}) classifies the eigenstates of ${\cal H}_0$ in three groups, according to the expectation value of $\Gamma$, that can be $0$, for bonding and antibonding states, and  $+1$ or $-1$ for sublattice polarized non-bonding states.

\begin{figure}
 \centering
    \includegraphics[width=0.45\textwidth]{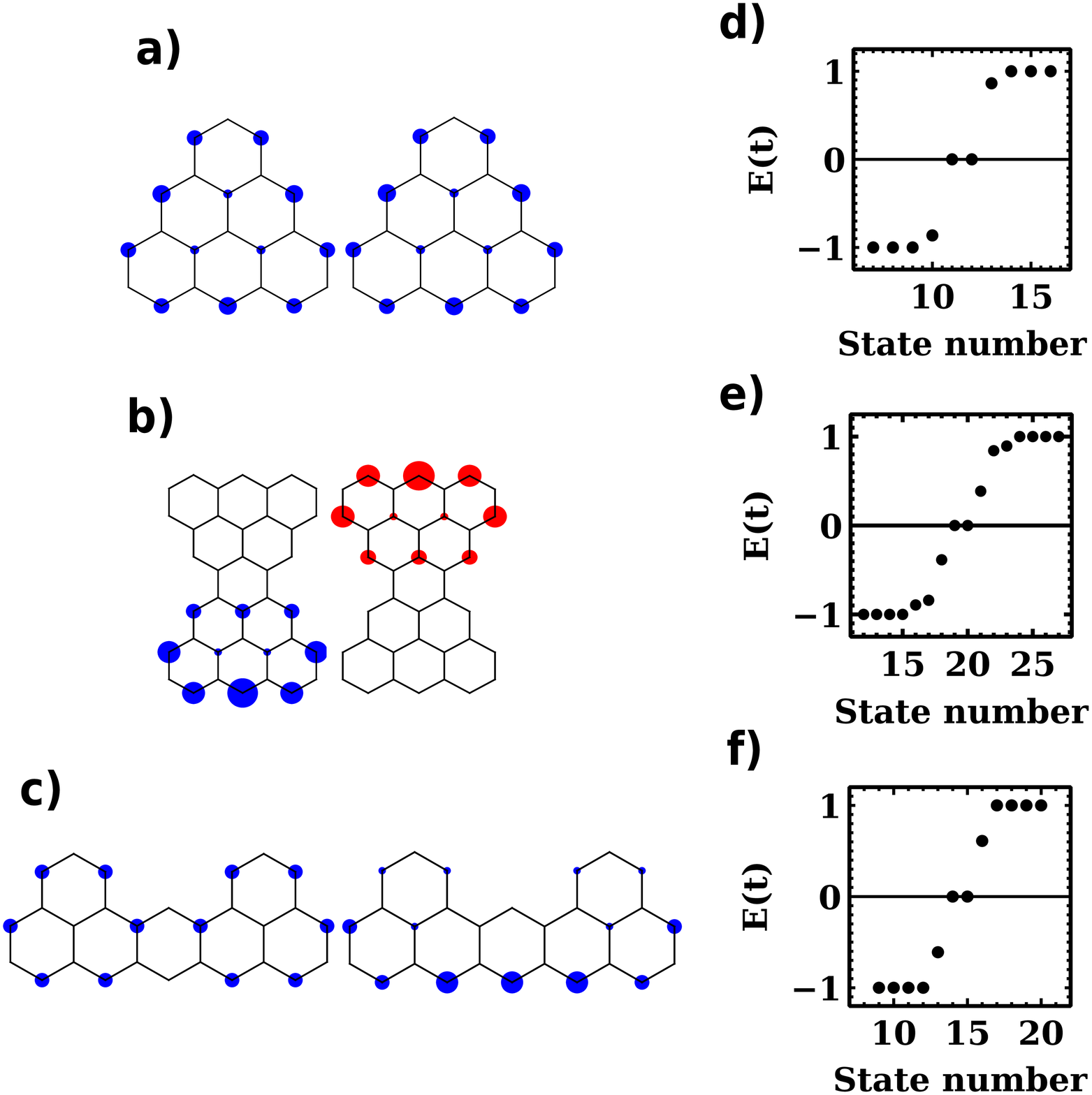}
\caption{ Non interacting zero modes wavefunction  (a), (b), (c)  and  single particle spectra (d, e, f) for three types of structure: triangulene (top), Clar's goblet (middle) and heptauthrene (bottom).  In the three cases the single-particle energy levels  has two $E=0$ states. The color (red, blue) represents  sublattice   and the area of the circles stands for $|\phi_1|^2$ and $|\phi_2|^2$.   }
\label{fig1}
\end{figure}

With this background, we now provide a straightforward demonstration of  the well known result\cite{LonguetHiggings, Borden77,sutherland86,inui1994} that for a given bipartite graph with $N_A-N_B\neq 0$ there are, {\em at  least}, $N_Z=|N_A-N_B|$ zero modes. For that matter, we first compute the trace of the $\Gamma$ operator, calculated in the atomic orbital basis:
\begin{equation}Tr \Gamma= \sum_i \langle i|\Gamma|i\rangle= N_A-N_B
\label{Tri}
\end{equation}
where $i$ labels the atomic sites. 
We now write down the trace in the basis of eigenstates of ${\cal H}_0$. We break  it down in terms of three types, namely,  states with $E_n\neq0$ that do not contribute to the trace,  and $A$ or $B$ zero modes, that contribute with $\pm 1$:
\begin{equation}
Tr \Gamma= \sum_n \langle \psi_n |\Gamma|\psi_n\rangle + \sum_A 1-\sum_B 1
\label{globalzero}
\end{equation}
Comparison of  Eq.(\ref{Tri}) and (\ref{globalzero}) establishes $N_{zA}-N_{zB}=N_A-N_B$, where $N_{zA/zB}$ is the number of zero modes localized in a given sublattice. 
  {\em Importantly},  Eq.(\ref{globalzero}) gives the minimum 
number of zero modes, as we can certainly  have $N_{zA}+N_{zB}> N_A-N_B$.  Thus, in this work we consider
systems, such as  the so-called triangulene and  heptauthrene (Fig.\ref{sublattices}a, c), that have sublattice imbalance of $2$ and  two zero modes  in the majority sublattice (Fig.\ref{fig1} a, c, d, f)  and  also systems with two zero modes and $N_A=N_B$ (Fig.\ref{fig1}b, e). 

We have found  two different  classes  of situations in which systems with $N_A=N_B$ have two zero modes:
\begin{itemize}
\item When there is a symmetry operation $\hat{\cal R}$  that conmutes both with ${\cal H}_0$ and $\Gamma$, such that the different irreducible representations of $\hat{\cal R}$ have a partial trace of $\Gamma$ different from zero\cite{Koshino}. This is the case of  $4n$ anulenes, such as cyclobutadiene.  Importantly, Jahn-Teller mechanism is operative here, and   a structural distortion of the PAH can  lift the degeneracy of the zero modes.  

\item When the structure is formed by fragments that, when considered separately,  have zero modes whose wavefunction have no weight on the junction region \cite{Borden77}.  This is the case
of  Clar's goblet\cite{clar72} (Fig.\ref{fig1}b, e). Here the zero modes are robust with respect to Jahn-Teller distorsions. 
\end{itemize} 
The fact that, for $N_A=N_B$,  the two zero modes are hosted in different sublattices has important consequences in the many-body Hamiltonian, as we discuss below. 

\section{PAH diradicals:  a Hubbard model description.}

\subsection{ Hubbard model in a minimal Active Space.}
The discussion of the effect of electron-electron interactions  starts with the description of the Hubbard model that, in the context of nanographenes,  has been shown\cite{Fernandez2007,Fernandez08}   to give results very similar to those of Density Functional Theory calculations. We start with the minimal Hilbert space for 
 two electrons in two molecular  orbitals,  $\phi_1$ and $\phi_2$, that are linear combination of the $\pi$ atomic orbitals of the PAH's C atoms.
 These molecular orbitals are obtained by diagonalization of a single-particle Hamiltonian, ${\cal H}_0$,  that describes first neighbor single orbital tight-binding model. The resulting spectra and molecular orbitals for a variety of nanographenes are shown in Fig.\ref{fig1}.  All of them have two   zero modes that are occupied with two electrons at half filling.  When interactions are ignored,  the two-fold orbital degeneracy of the  in-gap states results in a six-fold degeneracy of the multi-electronic  ground state,   that is lifted by Coulomb interactions.

When treated in the Hubbard approximation,  the interaction term is simply given by $U\sum_i n_{i\uparrow} n_{i\downarrow}$, where $n_{i\sigma}$ stands for the occupation of the atomic $\pi$ orbital with spin $\sigma$ at carbon $i$.    In contrast,  it can be seen that the same expression, projected over the two in-gap molecular orbitals results in a Hamiltonian with four different terms (see suppl. mat.):
\begin{eqnarray}
{\cal H}_{\cal U}&=& 
\left(\tilde{{\cal U}}_1-\frac{J}{4}\right)   n_{1\uparrow} n_{1\downarrow} +
\left(\tilde{{\cal U}}_2   -\frac{J}{4}\right)
  n_{2\uparrow} n_{2\downarrow}  
  -\nonumber\\
&-&  J \vec{S}_1\cdot\vec{S}_2+ {\cal V}_{\rm pair}+{\cal V}_{12}+ {\cal V}_{21}
\label{HAMIL}
\end{eqnarray}
where  $n_{\eta\sigma}= C^{\dagger}_{\eta\sigma}C_{\eta\sigma}$ is the occupation operator of the {\em molecular orbital} $\phi_\eta$,
$\vec{S}_\eta= \frac{1}{2} \sum_{\sigma,\sigma'} C^{\dagger}_{\eta\sigma}\vec{\tau}_{\sigma,\sigma'}C_{\eta\sigma'}$ is the spin operator associated to that orbital (see suppl. mat.) and $C^{\dagger}_{\eta\sigma}$ is the operator that creates and electron with spin $\sigma$ in the molecular orbital  $\phi_\eta$.

Hamiltonian  (\ref{HAMIL}) has four types of interactions.  First,  Hubbard-like terms, with energy ${\cal U}_{\eta}-\frac{J}{4}$, that describe  the  Coulomb energy penalty of double occupation of states $\phi_1$ and $\phi_2$. The energy scales ${\cal U}_{\eta}$ are given by: 
\begin{equation}
{\cal U}_{\eta}= U \sum_i |\phi_{\eta}(i)|^4
\label{U}
\end{equation}
where $\sum_i |\phi_{\eta}(i)|^4$ is the so-called inverse participation ratio and is a metric of the extension of the orbital\cite{Ortiz2018}.

The exchange integral is given by:
\begin{equation}
J= 2U \sum_i |\phi_{1}(i)|^2|\phi_{2}(i)|^2
\label{J}
\end{equation}
and is a metric of the {\em overlap} of the two zero modes.  It is apparent that $J=0$ for disjoint zero modes.

   Second, a ferromagnetic exchange term between the spins in orbitals (notice that $J\geq 0$). 
       The last two terms are 
   the  {\em density assisted} hopping terms, 
 \begin{eqnarray}
{\cal V}_{12}=\sum_{\sigma} n_{1\sigma}
\left(
t_{12}  C^{\dagger}_{1\overline{\sigma}} C_{2\overline{\sigma}}
+t_{12}^*  C^{\dagger}_{2\overline{\sigma}} C_{1\overline{\sigma}} \right)
\end{eqnarray}
and the pair hopping term:
\begin{eqnarray}
{\cal V}_{\rm pair}=
\Delta  C^{\dagger}_{1\uparrow} C^{\dagger}_{1\downarrow} C_{2\uparrow}C_{2\downarrow}
+ \Delta^*  C^{\dagger}_{2\uparrow} C^{\dagger}_{2\downarrow}  C_{1\uparrow}C_{1\downarrow} 
\end{eqnarray}

The Hubbard matrix elements that control these Coulomb assisted hoppings  are:
   \begin{equation}
t_{12}= U \sum_i |\phi_{1}(i)|^2 \phi_1(i)^* \phi_{2}(i)
\label{t12}
\end{equation}
and
   \begin{equation}
\Delta= U \sum_i (\phi_{1}(i)^*)^2 \phi_2(i)^2
\label{D}
\end{equation}
Both $t_{12}$ and $\Delta$ can be complex numbers. 

At this point, we make two crucial observations. First, since
$\phi_1$ and $\phi_2$  are degenerate,
 there is not a unique representation for them and the values of the Hubbard integrals ${\cal U}_{\eta}$, $J$, $\Delta$ and $t_{12}$ depend on our choice of representation.  
In the following we choose $\phi_1$ and $\phi_2$ so that they diagonalize a symmetry operator of the point group of the molecule. This permits to  obtain closed expressions for the spectrum of Hamiltonian (\ref{HAMIL}).  
To do so,  the first step is to  find a symmetry operator $\hat{\cal R}$ that conmutes with the single particle Hamiltonian, $[\hat{\cal R},{\cal H}_0]=0$ and with the sublattice operator $\Gamma$.  
We choose $\phi_1$ and $\phi_2$ such that:
\begin{eqnarray}
{\cal \hat{R}}\phi_{1,2} = \lambda_{1,2} \phi_{1,2}
 \end{eqnarray}
In table I the relevant symmetry operators and the eigenvalues $\lambda_{1,2}$ are listed  for the three systems of interest.  Symmetries include  a 120$^\circ$ rotation   in the case of triangulene, and reflection around the mirror symmetry  axis in the case of  Clar's goblet and heptauthrene.

Our second central observation is the fact that Hubbard integrals, defined in eqs.(\ref{U}, \ref{J}, \ref{t12}, \ref{D}) have to remain invariant under the symmetry operation. Hence,  if we replace $\phi_1$ and $\phi_2$  by
 ${\cal \hat{R}}\phi_1=\lambda_1 \phi_1 $ and ${\cal \hat{R}}\phi_2=\lambda_2 \phi_2$,  we have:
\begin{eqnarray}
{\cal U}_{\eta}&=& |\lambda_{\eta}|^4 {\cal U}_{\eta}\nonumber\\
J&=& |\lambda_{1}|^2 |\lambda_{2}|^2 J \nonumber\\
t_{12}&=& |\lambda_1|^2\lambda_{1}^* \lambda_{2} t_{12} \nonumber\\
\Delta&=& \left( \lambda_{1}^* \right)^2 \lambda_{2}^2 \Delta
\label{sym}
\end{eqnarray}

Thus,  the prefactor in the right hand side of Eq.(\ref{sym}) have to be identical to one, otherwise the corresponding Hubbard integral vanishes.  As it can be inferred from  table I,  the Hubbard integral  $t_{12}=0$   for  $C_3$ and heptauthrene systems, out of symmetry. Obviously, $t_{12}$ also vanishes for all  diradicals with disjoint zero modes.    When $t_{12}=0$, we obtain  analytical expressions for the six eigenvalues of Eq.(\ref{HAMIL}):
\begin{eqnarray}
E_T&=& -\frac{J}{4}\;   \nonumber \\
E_{S1}&=& +\frac{3J}{4}\;  \nonumber \\
E_{S2}&=&\overline{{\cal U}}-\frac{J}{4} -\sqrt{\Delta^2+\frac{1}{4}\left({\cal U}_1-{\cal U}_2\right)^2} \; 
  \nonumber \\
E_{S3}&=&\overline{{\cal U}}-\frac{J}{4} +\sqrt{\Delta^2+\frac{1}{4}\left({\cal U}_1-{\cal U}_2\right)^2} \;  
\end{eqnarray}
where $\overline{{\cal U}}\equiv \frac{1}{2} \left({\cal U}_1+{\cal U}_2\right)$.   The first line corresponds to the triplet.
The energy $E_{S1}$ corresponds to the open-shell singlet, and $E_{S2}$ and $E_{S3}$ to closed-shell singlets. These equations constitute our first important result. 

 

\begin{figure*}
 \centering
    \includegraphics[width=0.9 \textwidth]{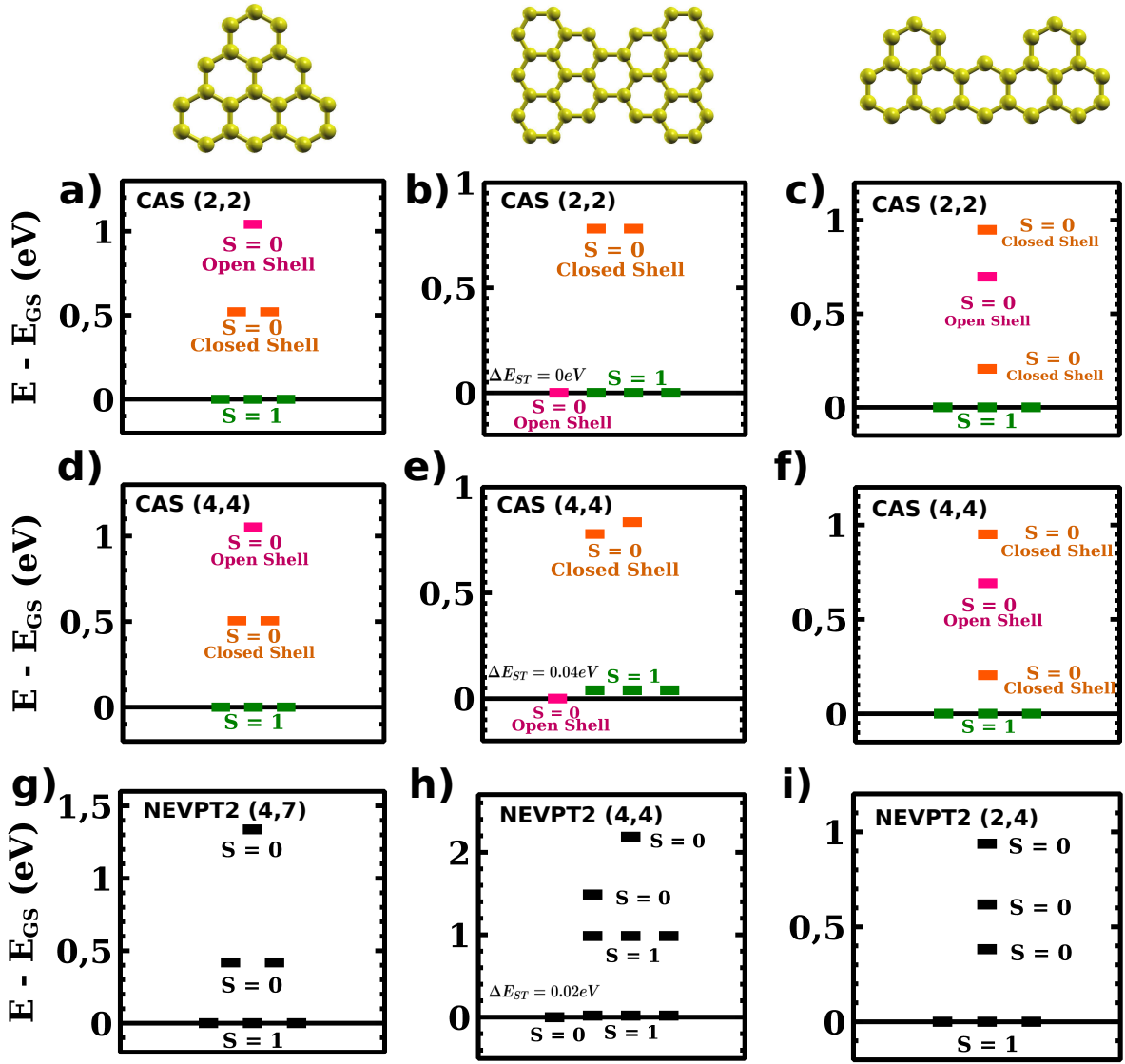}
\caption{Low energy excitation energy  spectra of various diradicals, including the effect of Coulomb interaction.  Top panels: spectra calculated using the analytic exact solution of the Hubbard model 
with an active space that only includes the zero modes, CAS(2,2). Middle panels: spectra calculated numerically with CAS(4,4) employing CI method. Lower panels: spectra calculated numerically with different active spaces employing the NEVPT2 method (see suppl. mat, section VI). For top and middle panels
we took 
$t = -2.7 eV$ and $U = |2t|$.}
\label{fig2}
\end{figure*}

\subsection{Hubbard model Reduced active space: theory}

In the following we apply this theory to three different types of diradicals, with and without sublattice imbalance and with different point group symmetries.  In all cases, 
we determine the open-/closed-shell of the multielectronic states by  the overlap of the  multielectronic wave function with the configurations with a well defined occupation of the single particle states, always taking the zero modes as eigenstates of a symmetry operator.

\subsubsection{ $C_3$ diradicals: Hubbard model.}
We consider first the case of  systems with  $N_A-N_B=2$ and $C_{3}$ symmetry, such as  triangulene\cite{Fernandez2007} and trimethylenemethane (see suppl. mat.).  In this case,   the  relevant symmetry is the in-plane 120$^{\circ}$  
rotation, $\hat {\cal R}(\frac{2\pi}{3})$.  The diagonal representation of $\hat {\cal R}(\frac{2\pi}{3})$ in the subspace of zero modes has two 
  eigenvalues  $e^{\pm i\epsilon}$ with $\epsilon= \frac{2\pi}{3}$. Therefore, given that the rotation matrix is real, we can easily see that $\phi_1(i)= \left(\phi_2(i)\right)^*$, i.e., overlap between zero modes is maximal.   From here, 
  we obtain ${\cal U}_{1}={\cal U}_2= \frac{J}{2}$, which implies that exchange is the dominant energy scale and   $\Delta=0$ and $t_{12}=0$.

 The resulting spectrum, shown in Fig.\ref{fig2}a, has a ground state triplet, in agreement with Lieb's theorem, with energy $E_T=-\frac{J}{4}$.  The three excited singlets  are arranged in a low energy closed-shell doublet, with energies $E_{S2}=E_{S3}=\frac{J}{4}$,  and a high energy open-shell singlet with energy $E_{S1}=3\frac{J}{4}$.       The  spectrum  of excited states  of the triangulene is quite peculiar for two reasons.  First, 
 the open-shell singlet is the highest  energy excitation,  within the manifold of states considered in this approximation.  This reflects  the large value of exchange
 that arises from the maximal overlap of the zero modes. Second, the closed-shell singlets are degenerate.

 \begin{table}[htp]
 \caption{Hubbard integrals for the structures from Fig.1}
 \begin{center}
    \begin{tabular}{| l | l | l | l | l | l | l | l | l | }
    \hline
    Structure & ${\cal U}_1/\overline{{\cal U}}$ & ${\cal U}_2/\overline{{\cal U}}$ &   $J/\overline{{\cal U}}$ & $t_{12}/\overline{{\cal U}}$ & $\Delta/\overline{{\cal U}}$ & $\overline{{\cal U}}(U)$ & $\lambda(\hat {\cal R})$\\ \hline
    Triangulene & 1 & 1  & 2 & 0  & 0 & 0.096 & $e^{\pm\frac{i2\pi}{3}}$\\ 
    Clar's goblet & 1 & 1  & 0 & 0   & 0 & 0.144 & $1$\\
    Heptauthrene & 1.220 & 0.779  & 1.209 & 0   & 0.604 & 0.106 & $\pm1$\\
    
    \hline
    \end{tabular}
\end{center}
 \end{table}

\subsubsection{ Diradicals with $N_A-N_B=2$ and reflection  symmetry.}
 We now discuss the case of PAHs with a different point group.   We consider the heptauthrene diradical, that has  a reflection symmetry that preserves sublattice.  The eigenvalues of this symmetry operator are $\pm 1$; thus, its application leaves all the Hubbard integrals unaffected except  for $t_{12}$ and $t_{21}$ that  change sign and, therefore,  must vanish identically. Since  the eigenvalues of the symmetry operator are  real,
 the states $\phi_1$ and $\phi_2$ are also real,  which automatically entails $J=2\Delta$.  However,     
 we have ${\cal U}_1\neq{\cal U}_2$.  The resulting many-body spectrum, within the reduced active space,  is shown in Fig.\ref{fig2}c. The ground state has $S=1$, complying with the Lieb's theorem. The excited states follow a more  conventional arrangement,  with three non-degenerate singlets, with the  open-shell singlet  in between the two closed-shell singlets.
 
\subsubsection{ $N_A=N_B$ diradicals.}
 We now apply Hamiltonian (\ref{HAMIL}) to the case of  diradicals with $N_A=N_B$ such as the Clar's  goblet \cite{clar72,Yazyev09, Fajtlowicz2005OnMM} and the cyclobutadiene\cite{Koshino, SchumannR}  (see suppl. mat.). 
The single-particle zero modes  of diradicals with $N_A=N_B$ are disjoint, i.e.,  they are located in different atoms\cite{Kenneth}.  As a result,  the only non-zero energy scales of  the restricted many-body Hamiltonian (\ref{HAMIL})   are ${\cal U}_{1,2}$.   
 The resulting interacting spectrum (Fig.\ref{fig2}b), in the minimal active space,  presents a quartet ground state, formed by  the open-shell  $S=0$ and $S=1$ states, and two closed-shell excited $S=0$ states.    Extension of the Hilbert space is thus necessary to resolve the singlet-triplet energy difference, that must favour a $S=0$ ground state to comply with Lieb's theorem\cite{Lieb89}.
 
\section{ Hubbard model calculations with an extended Hilbert space.}
The description of the diradicals using a restricted Hilbert space  sometimes gives  spectra with peculiar degeneracies, such as the  first excited state  of the $C_3$ diradicals or the ground state degeneracy of the $N_A=N_B$ diradicals.  We now explore whether these degeneracies are artefacts of the truncated Hilbert space, or they are byproducts of symmetry.  For that matter, we have carried out exact diagonalizations  using a configuration interaction  method for the Hubbard model,  extending the active space to include valence and conduction states.  We denote a complete active space (CAS) 
with $\cal N$ electrons and  $\cal M$ molecular orbitals, each with a two-fold spin degeneracy, as CAS($\cal N$, $\cal M$).  For ${\cal N}={\cal M}=4$ we have a total of    70  many-body states.

Our results for CAS(4,4) for the three systems of interest are shown in the second row of Fig.(\ref{fig2}).
In the case of triangulene, the main peculiar properties of the spectrum obtained in the minimal model are preserved here.  This is also  the case of trimethylenemethane (see suppl. mat.),  for which the CAS(4,4) calculation covers the complete Hilbert space. We thus conclude that the peculiar symmetry obtained in the analytical CAS(2,2) model is not a specific feature of a truncated active space. Our results for heptauthrene, using CAS(4,4) are also in line with the analytical model. 

Unlike the analytical model, our  CAS(4,4) calculations give  non-degenerate  $S=0$ ground state for the 
 diradicals with $N_A=N_B$, such as the Clar's goblet and the cyclobutadiene. This complies with Lieb's theorem\cite{Lieb89}.      The lowest excited state is a triplet.   We have verified (see suppl. mat.) that the singlet-triplet splitting, scales  as
$E_{S=1}-E_{S=0}\propto  U^2/t>0$.  This scaling indicates that the antiferromagnetic exchange is driven by virtual excursions to excited states, with electrons in the conduction states and/or holes in the valence states,   driven by the Hubbard term. 
 It must be noted that this is different from the usual kinetic exchange\cite{Anderson59} scaling $t^2/U$, that has also been discussed for 
 the case of  in-gap zero modes  hybridized by  hopping in nanographenes\cite{Ortiz2018}.

\section{ Ab initio  calculations.}
At this point we further explore to which point the predictions of the Hubbard model are different from more sophisticated yet costly
ab initio calculations including nuclear coordinates relaxation, more atomic orbitals per atom,  H atoms passivating edge C atoms, and 
long-range Coulomb interactions. The structures and molecular orbital basis set used to carry out the multi-configurational approaches are obtained 
using a Density-Functional Theory calculation based with the hybrid exchange-correlation functional B3LYP (see suppl. mat.).  
 
We have carried out ab initio calculations for three different diradicals: triangulene, Clar's Goblet and heptauthrene (see Fig.\ref{fig2}). In all three cases, the ground state and first excited state have the same spin than anticipated in the previous sections. 
In the case of triangulene, the symmetry of the spectrum also comes in the 3-2-1 sequence obtained using the Hubbard model. The relative value of the  energy difference between 
triplet and open-shell singlet is also  in line with the Hubbard model: largest for triangulene, and much smaller for the Clar's goblet.

In the case of  the Clar's goblet, the ab initio calculations predict  a second excited triplet  with energy smaller than that of the   two singlets, in contrast with the Hubbard model predictions.  This discrepancy is probably related to the 
underestimation of the energy overhead associated to the double occupancy of a molecular orbital in the closed-shell configuration  when electronic repulsion is treated in the Hubbard approximation. 
As a result, the Hubbard model underestimates the energy of the  two closed-shell singlets, compared to the second open-shell triplet.  Barring this discrepancy in the higher order excited states of the Clar's goblet, 
the main features of the Hubbard model are confirmed with the more sophisticated, but much more computationally  expensive, ab initio methods.

\section{ Exchange rules.}
We are now in position to conclude with a number of  rules for the exchange interactions in PAHs diradicals, based on the analysis of the results obtained with various methods:
\begin{enumerate}
\item Lieb's rule:  the spin of the ground state, $S$, is determined by the  sublattice imbalance, $|N_A-N_B|= 2S$.

\item The  spin and orbital degeneracies of the  six lowest energy states is predicted, qualitatively,  by Hamiltonian (\ref{HAMIL}).
The eigenvalues   energy scales are governed by ${\cal U}_{1,2}$, $J$, $\Delta$ that, in turn, are strongly 
conditioned by the point symmetry group of the PAH diradical. 

\item Ferromagnetic exchange $J$ is maximal for $C_3$ diradicals, on account of  the maximal overlap of the zero modes.   The $C_3$ symmetry also imposes that  the  lowest energy excited state turns out to be given by a degenerate pair of closed-shell states.  

\item Ferromagnetic exchange $J$ is minimal for $N_A=N_B$ diradicals, on account of the disjoint nature of their zero modes.

\end{enumerate}

We acknowledge J. L. Lado for fruitful discussions.
J. F.-R.  and R. O. acknowledge financial support from MINECO-Spain (Grant No. MAT2016-78625-C2) and 
from  the Portuguese ``Funda\c{c}\~ao para a Ci\^encia e a Tecnologia''  (FCT) for the project  P2020-PTDC/FIS-NAN/4662/2014.
J.F.-R., M. M.-F.  and N. G.-M.  acknowledge support from  the P2020-PTDC/FIS-NAN/3668/2014.
J. F.-R. acknowledges support from 
 UTAPEXPL/NTec/0046/2017 projects, as well as Generalitat
Valenciana funding (Prometeo2017/139).
 R. O. and J. C. S.-G. acknowledge ACIF/2018/175 (Generalitat Valenciana and Fondo Social Europeo).
M. M.-F. and R. B. would like to acknowledge the Portuguese ``Funda\c{c}\~ao para a Ci\^encia e a Tecnologia''  (FCT) for the project IF/00894/2015 and FCT Ref. UID/CTM/50011/2019 for CICECO - Aveiro Institute of Materials. 
This project has received funding from the European Union's Horizon 2020 research and innovation programme under grant agreement No 664878.


\bibliographystyle{naturemag}
\bibliography{biblio}{}

\begin{thebibliography}{10}
\expandafter\ifx\csname url\endcsname\relax
  \def\url#1{\texttt{#1}}\fi
\expandafter\ifx\csname urlprefix\endcsname\relax\def\urlprefix{URL }\fi
\providecommand{\bibinfo}[2]{#2}
\providecommand{\eprint}[2][]{\url{#2}}

\bibitem{goodenough63}
\bibinfo{author}{Goodenough, J.~B.}
\newblock \emph{\bibinfo{title}{Magnetism and chemical bond}},
  vol.~\bibinfo{volume}{1} (\bibinfo{publisher}{Interscience Publ.},
  \bibinfo{year}{1963}).

\bibitem{kanamori59}
\bibinfo{author}{Kanamori, J.}
\newblock \bibinfo{title}{Superexchange interaction and symmetry properties of
  electron orbitals}.
\newblock \emph{\bibinfo{journal}{J. Phys. Chem. Solids}}
  \textbf{\bibinfo{volume}{10}}, \bibinfo{pages}{87--98}
  (\bibinfo{year}{1959}).

\bibitem{Schlenk}
\bibinfo{author}{Schlenk, W.} \& \bibinfo{author}{Brauns, M.}
\newblock \bibinfo{title}{Zur frage der metachinoide}.
\newblock \emph{\bibinfo{journal}{Ber. Dtsch. Chem. Ges.}}
  \textbf{\bibinfo{volume}{48}}, \bibinfo{pages}{661--669}
  (\bibinfo{year}{1915}).

\bibitem{Borden77}
\bibinfo{author}{Borden, W.~T.} \& \bibinfo{author}{Davidson, E.~R.}
\newblock \bibinfo{title}{Effects of electron repulsion in conjugated
  hydrocarbon diradicals}.
\newblock \emph{\bibinfo{journal}{JACS}} \textbf{\bibinfo{volume}{99}},
  \bibinfo{pages}{4587--4594} (\bibinfo{year}{1977}).

\bibitem{rajcareview}
\bibinfo{author}{Rajca, A.}
\newblock \bibinfo{title}{Organic diradicals and polyradicals: From spin
  coupling to magnetism?}
\newblock \emph{\bibinfo{journal}{Chem. Rev.}} \textbf{\bibinfo{volume}{94}},
  \bibinfo{pages}{871--893} (\bibinfo{year}{1994}).

\bibitem{moritaReview}
\bibinfo{author}{Morita, Y.}, \bibinfo{author}{Suzuki, S.} \&
  \bibinfo{author}{Takui, T.}
\newblock \bibinfo{title}{Synthetic organic spin chemistry for structurally
  well-defined open-shell graphene fragments}.
\newblock \emph{\bibinfo{journal}{Nat. Chem.}} \textbf{\bibinfo{volume}{3}},
  \bibinfo{pages}{197--204} (\bibinfo{year}{2011}).

\bibitem{pavlivcek2017}
\bibinfo{author}{Pavliček, N.} \emph{et~al.}
\newblock \bibinfo{title}{Synthesis and characterization of triangulene}.
\newblock \emph{\bibinfo{journal}{Nat. Nano.}} \textbf{\bibinfo{volume}{18}},
  \bibinfo{pages}{308–311} (\bibinfo{year}{2017}).

\bibitem{wang2016}
\bibinfo{author}{Wang, S.} \emph{et~al.}
\newblock \bibinfo{title}{Giant edge state splitting at atomically precise
  graphene zigzag edges}.
\newblock \emph{\bibinfo{journal}{Nat. Comm.}} \textbf{\bibinfo{volume}{7}}
  (\bibinfo{year}{2016}).

\bibitem{NachoNat}
\bibinfo{author}{Li, J.} \emph{et~al.}
\newblock \bibinfo{title}{Single spin localization and manipulation in graphene
  open-shell nanostructures}.
\newblock \emph{\bibinfo{journal}{Nat. Comm.}} \textbf{\bibinfo{volume}{10}}
  (\bibinfo{year}{2019}).

\bibitem{Yoneda2009}
\bibinfo{author}{Yoneda, K.} \emph{et~al.}
\newblock \bibinfo{title}{Third-order nonlinear optical properties of trigonal,
  rhombic and bow-tie graphene nanoflakes with strong structural dependence of
  diradical character}.
\newblock \emph{\bibinfo{journal}{Chem. Phys. Chem.}}
  \textbf{\bibinfo{volume}{480}}, \bibinfo{pages}{278--283}
  (\bibinfo{year}{2009}).

\bibitem{melle2015}
\bibinfo{author}{Melle-Franco, M.}
\newblock \bibinfo{title}{Uthrene, a radically new molecule?}
\newblock \emph{\bibinfo{journal}{Chem. Comm.}} \textbf{\bibinfo{volume}{51}},
  \bibinfo{pages}{5387--5390} (\bibinfo{year}{2015}).

\bibitem{gryn'ova2015}
\bibinfo{author}{Gryn'ova, G.}, \bibinfo{author}{L.~Coote, M.} \&
  \bibinfo{author}{Corminboeuf, C.}
\newblock \bibinfo{title}{Theory and practice of uncommon molecular electronic
  configurations}.
\newblock \emph{\bibinfo{journal}{WIREs Comput. Mol. Sci.}}
  \textbf{\bibinfo{volume}{5}}, \bibinfo{pages}{440--459}
  (\bibinfo{year}{2015}).

\bibitem{Lischka2016}
\bibinfo{author}{Das, A.}, \bibinfo{author}{M\"uller, T.},
  \bibinfo{author}{Plasser, F.} \& \bibinfo{author}{Lischka, H.}
\newblock \bibinfo{title}{Polyradical character of triangular non-kekulé
  structures, zethrenes, p-quinodimethane-linked bisphenalenyl, and the clar
  goblet in comparison: An extended multireference study}.
\newblock \emph{\bibinfo{journal}{J. Phys. Chem. A}}
  \textbf{\bibinfo{volume}{120}}, \bibinfo{pages}{1625--1636}
  (\bibinfo{year}{2016}).

\bibitem{Sandoval2019}
\bibinfo{author}{Sandoval-Salinas, M.~E.}, \bibinfo{author}{Carreras, A.} \&
  \bibinfo{author}{Casanova, D.}
\newblock \bibinfo{title}{Triangular graphene nanofragments: open-shell
  character and doping}.
\newblock \emph{\bibinfo{journal}{PCCP}} \textbf{\bibinfo{volume}{21}},
  \bibinfo{pages}{9069--9076} (\bibinfo{year}{2019}).

\bibitem{Yoneda2011}
\bibinfo{author}{Yoneda, K.} \emph{et~al.}
\newblock \bibinfo{title}{Open-shell characters and second
  hyperpolarizabilities of one-dimensional graphene nanoflakes composed of
  trigonal graphene units}.
\newblock \emph{\bibinfo{journal}{Chem. Phys. Chem.}}
  \textbf{\bibinfo{volume}{12}}, \bibinfo{pages}{1697--1707}
  (\bibinfo{year}{2011}).

\bibitem{Yazyev09}
\bibinfo{author}{Wang, W.~L.}, \bibinfo{author}{Yazyev, O.~V.},
  \bibinfo{author}{Meng, S.} \& \bibinfo{author}{Kaxiras, E.}
\newblock \bibinfo{title}{Topological frustration in graphene nanoflakes:
  Magnetic order and spin logic devices}.
\newblock \emph{\bibinfo{journal}{Phys. Rev. Lett.}}
  \textbf{\bibinfo{volume}{102}}, \bibinfo{pages}{157201}
  (\bibinfo{year}{2009}).

\bibitem{Sheng2013}
\bibinfo{author}{Sheng, W.}, \bibinfo{author}{Sun, M.} \&
  \bibinfo{author}{Zhou, A.}
\newblock \bibinfo{title}{Violation of hund's rule and quenching of long-range
  electron-electron interactions in graphene nanoflakes}.
\newblock \emph{\bibinfo{journal}{Phys. Rev. B}} \textbf{\bibinfo{volume}{88}},
  \bibinfo{pages}{085432} (\bibinfo{year}{2013}).

\bibitem{Trinquier2015}
\bibinfo{author}{Trinquier, G.} \& \bibinfo{author}{Malrieu, J.-P.}
\newblock \bibinfo{title}{Kekulé versus lewis: When aromaticity prevents
  electron pairing and imposes polyradical character}.
\newblock \emph{\bibinfo{journal}{Chem.-A Eur. J.}}
  \textbf{\bibinfo{volume}{21}}, \bibinfo{pages}{814--828}
  (\bibinfo{year}{2015}).

\bibitem{Malrieu2016}
\bibinfo{author}{Malrieu, J.-P.} \& \bibinfo{author}{Trinquier, G.}
\newblock \bibinfo{title}{Can a topological approach predict spin-symmetry
  breaking in conjugated hydrocarbons?}
\newblock \emph{\bibinfo{journal}{J. of Chem. Phys. A}}
  \textbf{\bibinfo{volume}{120}}, \bibinfo{pages}{9564--6578}
  (\bibinfo{year}{2016}).

\bibitem{Trinquier2018}
\bibinfo{author}{Trinquier, G.} \& \bibinfo{author}{Malrieu, J.-P.}
\newblock \bibinfo{title}{Predicting the open-shell character of polycyclic
  hydrocarbons in terms of clar sextets}.
\newblock \emph{\bibinfo{journal}{J. Phys. Chem. A}}
  \textbf{\bibinfo{volume}{122}}, \bibinfo{pages}{1088--1103}
  (\bibinfo{year}{2018}).

\bibitem{Pogodin1985}
\bibinfo{author}{Pogodin, S.} \& \bibinfo{author}{Agranat, I.}
\newblock \bibinfo{title}{Clar goblet and related non-kekulé benzenoid lpahs. a
  theoretical study}.
\newblock \emph{\bibinfo{journal}{J. of Org. Chem.}}
  \textbf{\bibinfo{volume}{23}}, \bibinfo{pages}{698--704}
  (\bibinfo{year}{1985}).

\bibitem{ovchinnikov1978}
\bibinfo{author}{Ovchinnikov, A.~A.}
\newblock \bibinfo{title}{Multiplicity of the ground state of large alternant
  organic molecules with conjugated bonds}.
\newblock \emph{\bibinfo{journal}{Theor. Chim. Acta}}
  \textbf{\bibinfo{volume}{47}}, \bibinfo{pages}{297--304}
  (\bibinfo{year}{1978}).

\bibitem{Lieb89}
\bibinfo{author}{Lieb, E.~H.}
\newblock \bibinfo{title}{Two theorems on the hubbard model}.
\newblock \emph{\bibinfo{journal}{Phys. Rev. Lett.}}
  \textbf{\bibinfo{volume}{62}}, \bibinfo{pages}{1201--1204}
  (\bibinfo{year}{1989}).

\bibitem{sutherland86}
\bibinfo{author}{Sutherland, B.}
\newblock \bibinfo{title}{Localization of electronic wave functions due to
  local topology}.
\newblock \emph{\bibinfo{journal}{Phys. Rev. B}} \textbf{\bibinfo{volume}{34}},
  \bibinfo{pages}{5208--5211} (\bibinfo{year}{1986}).

\bibitem{cusinato2018}
\bibinfo{author}{Cusinato, L.}, \bibinfo{author}{Evangelisti, S.},
  \bibinfo{author}{Leininger, T.} \& \bibinfo{author}{Monari, A.}
\newblock \bibinfo{title}{The electronic structure of graphene nanoislands: A
  cas-scf and nevpt2 study}.
\newblock \emph{\bibinfo{journal}{Adv. Cond. Matt. Phys.}}
  \textbf{\bibinfo{volume}{2018}}, \bibinfo{pages}{14} (\bibinfo{year}{2018}).

\bibitem{longuet50}
\bibinfo{author}{Longuet-Higgins, H.}
\newblock \bibinfo{title}{Some studies in molecular orbital theory i. resonance
  structures and molecular orbitals in unsaturated hydrocarbons}.
\newblock \emph{\bibinfo{journal}{J. Chem. Phys.}}
  \textbf{\bibinfo{volume}{18}}, \bibinfo{pages}{265--274}
  (\bibinfo{year}{1950}).

\bibitem{inui1994}
\bibinfo{author}{Inui, M.}, \bibinfo{author}{Trugman, S.} \&
  \bibinfo{author}{Abrahams, E.}
\newblock \bibinfo{title}{Unusual properties of midband states in systems with
  off-diagonal disorder}.
\newblock \emph{\bibinfo{journal}{Phys. Rev. B}} \textbf{\bibinfo{volume}{49}},
  \bibinfo{pages}{3190} (\bibinfo{year}{1994}).

\bibitem{Fajtlowicz2005OnMM}
\bibinfo{author}{Fajtlowicz, S.}, \bibinfo{author}{John, P.~E.} \&
  \bibinfo{author}{Sachsb, H.}
\newblock \bibinfo{title}{On maximum matchings and eigenvalues of benzenoid
  graphs *}.
\newblock \emph{\bibinfo{journal}{CCACCA}} \textbf{\bibinfo{volume}{78}},
  \bibinfo{pages}{195--201} (\bibinfo{year}{2005}).

\bibitem{Fernandez2007}
\bibinfo{author}{Fern\'andez-Rossier, J.} \& \bibinfo{author}{Palacios, J.~J.}
\newblock \bibinfo{title}{Magnetism in graphene nanoislands}.
\newblock \emph{\bibinfo{journal}{Phys. Rev. Lett.}}
  \textbf{\bibinfo{volume}{99}}, \bibinfo{pages}{177204}
  (\bibinfo{year}{2007}).

\bibitem{weikprb2016}
\bibinfo{author}{Weik, N.}, \bibinfo{author}{Schindler, J.},
  \bibinfo{author}{Bera, S.}, \bibinfo{author}{Solomon, G.~C.} \&
  \bibinfo{author}{Evers, F.}
\newblock \bibinfo{title}{Graphene with vacancies: Supernumerary zero modes}.
\newblock \emph{\bibinfo{journal}{Phys. Rev. B}} \textbf{\bibinfo{volume}{94}},
  \bibinfo{pages}{064204} (\bibinfo{year}{2016}).

\bibitem{Ryu02}
\bibinfo{author}{Ryu, S.} \& \bibinfo{author}{Hatsugai, Y.}
\newblock \bibinfo{title}{Topological origin of zero-energy edge states in
  particle-hole symmetric systems}.
\newblock \emph{\bibinfo{journal}{Phys. Rev. Lett.}}
  \textbf{\bibinfo{volume}{89}}, \bibinfo{pages}{077002}
  (\bibinfo{year}{2002}).
\newblock
  \urlprefix\url{https://link.aps.org/doi/10.1103/PhysRevLett.89.077002}.

\bibitem{delplace2011}
\bibinfo{author}{Delplace, P.}, \bibinfo{author}{Ullmo, D.} \&
  \bibinfo{author}{Montambaux, G.}
\newblock \bibinfo{title}{Zak phase and the existence of edge states in
  graphene}.
\newblock \emph{\bibinfo{journal}{Physical Review B}}
  \textbf{\bibinfo{volume}{84}}, \bibinfo{pages}{195452}
  (\bibinfo{year}{2011}).

\bibitem{LonguetHiggings}
\bibinfo{author}{Longuet‐Higgins, H.~C.}
\newblock \bibinfo{title}{Some studies in molecular orbital theory i. resonance
  structures and molecular orbitals in unsaturated hydrocarbons}.
\newblock \emph{\bibinfo{journal}{J. Chem. Phys.}}
  \textbf{\bibinfo{volume}{18}}, \bibinfo{pages}{265--274}
  (\bibinfo{year}{1950}).

\bibitem{Koshino}
\bibinfo{author}{Koshino, M.}, \bibinfo{author}{Morimoto, T.} \&
  \bibinfo{author}{Sato, M.}
\newblock \bibinfo{title}{Topological zero modes and dirac points protected by
  spatial symmetry and chiral symmetry}.
\newblock \emph{\bibinfo{journal}{Phys. Rev. B}} \textbf{\bibinfo{volume}{90}},
  \bibinfo{pages}{115207} (\bibinfo{year}{2014}).

\bibitem{clar72}
\bibinfo{author}{Clar, E.} \& \bibinfo{author}{Mackay, C.}
\newblock \bibinfo{title}{Circobiphenyl and the attempted synthesis of 1: 14,
  3: 4, 7: 8, 10: 11-tetrabenzoperopyrene}.
\newblock \emph{\bibinfo{journal}{Tetrahedron}} \textbf{\bibinfo{volume}{28}},
  \bibinfo{pages}{6041--6047} (\bibinfo{year}{1972}).

\bibitem{Fernandez08}
\bibinfo{author}{Fern\'andez-Rossier, J.}
\newblock \bibinfo{title}{Prediction of hidden multiferroic order in graphene
  zigzag ribbons}.
\newblock \emph{\bibinfo{journal}{Phys. Rev. B}} \textbf{\bibinfo{volume}{77}},
  \bibinfo{pages}{075430} (\bibinfo{year}{2008}).
\newblock \urlprefix\url{https://link.aps.org/doi/10.1103/PhysRevB.77.075430}.

\bibitem{Ortiz2018}
\bibinfo{author}{Ortiz, R.}, \bibinfo{author}{Garc\'{\i}a-Mart\'{\i}nez,
  N.~A.}, \bibinfo{author}{Lado, J.~L.} \&
  \bibinfo{author}{Fern\'andez-Rossier, J.}
\newblock \bibinfo{title}{Electrical spin manipulation in graphene
  nanostructures}.
\newblock \emph{\bibinfo{journal}{Phys. Rev. B}} \textbf{\bibinfo{volume}{97}},
  \bibinfo{pages}{195425} (\bibinfo{year}{2018}).

\bibitem{SchumannR}
\bibinfo{author}{Schumann, R.}
\newblock \bibinfo{title}{Thermodynamics of a 4-site hubbard model by
  analytical diagonalization}.
\newblock \emph{\bibinfo{journal}{Annalen der Physik}}
  \textbf{\bibinfo{volume}{11}}, \bibinfo{pages}{49--88}.

\bibitem{Kenneth}
\bibinfo{author}{Pozun, Z.~D.}, \bibinfo{author}{Su, X.} \&
  \bibinfo{author}{Jordan, K.~D.}
\newblock \bibinfo{title}{Establishing the ground state of the disjoint
  diradical tetramethyleneethane with quantum monte carlo}.
\newblock \emph{\bibinfo{journal}{J. Am. Chem. Soc.}}
  \textbf{\bibinfo{volume}{135}}, \bibinfo{pages}{13862--13869}
  (\bibinfo{year}{2013}).

\bibitem{Anderson59}
\bibinfo{author}{Anderson, P.~W.}
\newblock \bibinfo{title}{New approach to the theory of superexchange
  interactions}.
\newblock \emph{\bibinfo{journal}{Phys. Rev.}} \textbf{\bibinfo{volume}{115}},
  \bibinfo{pages}{2--13} (\bibinfo{year}{1959}).
\newblock \urlprefix\url{https://link.aps.org/doi/10.1103/PhysRev.115.2}.

\end{thebibliography}

\end{document}